\documentclass[12pt,preprint]{aastex}
\newcommand{\kms}{${\rm km~s}^{-1}$}
\shorttitle{Velocity Groups in the Pisces Ovedensity}
\shortauthors{Sesar et al.}

\begin{document}

\renewcommand{\thefootnote}{\fnsymbol{footnote}}
\title{Halo Velocity Groups in the Pisces Overdensity}
\renewcommand{\thefootnote}{\arabic{footnote}}

\author{
Branimir Sesar\altaffilmark{\ref{Washington}},
A.~Katherina Vivas\altaffilmark{\ref{CIDA}},
Sonia Duffau\altaffilmark{\ref{Heidelberg}},
\v{Z}eljko Ivezi\'{c}\altaffilmark{\ref{Washington}}
}
\altaffiltext{1}{University of Washington, Department of Astronomy, P.O.~Box
                 351580, Seattle, WA 98195-1580;
                 \{bsesar,~zi\}@u.washington.edu\label{Washington}}
\altaffiltext{2}{Centro de Investigaciones de Astronom\'{\i}a (CIDA), Apartado
                 Postal 264, M\'erida 5101-A, Venezuela; akvivas@cida.ve\label{CIDA}}
\altaffiltext{3}{Astronomisches Rechen-Institut, Zentrum f\"{u}r Astronomie der Universit\"{a}t Heidelberg,
                 M\"{o}nchhofstra\ss{}e 12-14, D-69120 Heidelberg, Germany;
                 sonia.duffau@gmail.com\label{Heidelberg}}

\begin{abstract}
We report spectroscopic observations with the Gemini South Telescope of 5 faint
(V$\sim 20$) RR Lyrae stars associated with the Pisces overdensity. At a
heliocentric and galactocentric distance of $\sim80$ kpc, this is the most
distant substructure in the Galactic halo known to date. We combined our
observations with literature data and confirmed that the substructure is
composed of two different kinematic groups. The main group contains 8 stars and
has $\langle V_{gsr}\rangle=50$ \kms, while the second group contains four stars
at a velocity of $\langle V_{gsr}\rangle=-52$ \kms, where $V_{gsr}$ is the
radial velocity in the galactocentric standard of rest. The metallicity
distribution of RR Lyrae stars in the Pisces overdensity is centered on
$[Fe/H]=-1.5$ dex and has a width of 0.3 dex. The new data allowed us to
establish that both groups are spatially extended making it very unlikely that
they are bound systems, and are more likely to be debris of a tidally disrupted
galaxy or galaxies. Due to small sky coverage, it is still unclear whether these
groups have the same or different progenitors.
\end{abstract}

\keywords{stars: variables: other --- Galaxy: halo ---
Galaxy: kinematics  --- Galaxy: structure}

\section{Introduction \label{introduction}}

In the last decade, abundant observational evidence has been accumulated in
favor of a hierarchical formation scenario of the halo of the Milky Way (see a
review by \citealt{helmi08}). The substructures in the halo have been observed
both as overdensities in space and as moving groups. The tidal streams of the
disrupting Sagittarius dwarf spheroidal (dSph) galaxy are the best example of
such substructures, with the streams wrapping around most of the sky
\citep{ivezic00,yanny00,vivas01,ibata02,majewski03}. Other known substructures
include the Virgo Stellar Stream \citep{duffau06,prior09} and several other
small structures associated with the Virgo overdensity
\citep{newberg07,juric08,vivas08}, the Monoceros structure
\citep{ibata03,yanny03,ivezic08}, the Triangulum-Andromeda overdensity
\citep{rocha04}, the Hercules-Aquila cloud \citep{belokurov07a}, the Orphan
stream \citep{belokurov07b}, and other smaller overdensities
\citep{clewley06,starkenburg09} and streams \citep{grillmair09,schlaufman09}.
Using Sloan Digital Sky Survey (SDSS; \citealt{york00}) multi-epoch data in the
stripe 82 region ($308\arcdeg<R.A. <60\arcdeg$, $|Dec|<1.25\arcdeg$),
\citet{sesar07} found an overdensity of RR Lyrae stars in the halo at a
heliocentric and galactocentric distance of $\sim80$ kpc and
$R.A.\sim354\arcdeg$ (termed ``Structure J'', see Figures 10 and 11 from
\citealt{sesar10}). This structure was subsequently renamed the
``Pisces overdensity'' by \citet{watkins09} after the constellation where it is
located.

Convincing evidence that the Pisces overdensity is indeed a coherent structure,
and not a random superposition of stars, was recently presented by
\citet{kollmeier09}. They measured radial velocities of 8 RR Lyrae stars located
in the Pisces overdensity and found that 5 of them have a narrow velocity
distribution centered on $\sim -75$ {\kms } in the heliocentric rest frame. The
authors found hints of a second velocity peak in their data, containing two or
maybe three stars, at a lower velocity ($<-150$ {\kms } in the heliocentric
rest frame). \citet{kollmeier09} interpreted the main peak as a dwarf galaxy or
a disrupting dwarf galaxy, and did not consider the second peak as a real group
due to a small number of stars. In this paper, we expand on their work by
reporting radial velocities and metallicities for 5 RR Lyrae stars in the Pisces
overdensity, four of which were not observed by \citet{kollmeier09}. By
combining these two data sets, we are now in position to confirm the existence
of a second peak in the velocity distribution, and are able to better explore
the spatial extent and metallicity distributions of these two substructures
present in the Pisces overdensity.

\section{Observations}

\subsection{Target Selection}

\begin{figure}
\epsscale{0.9}
\plotone{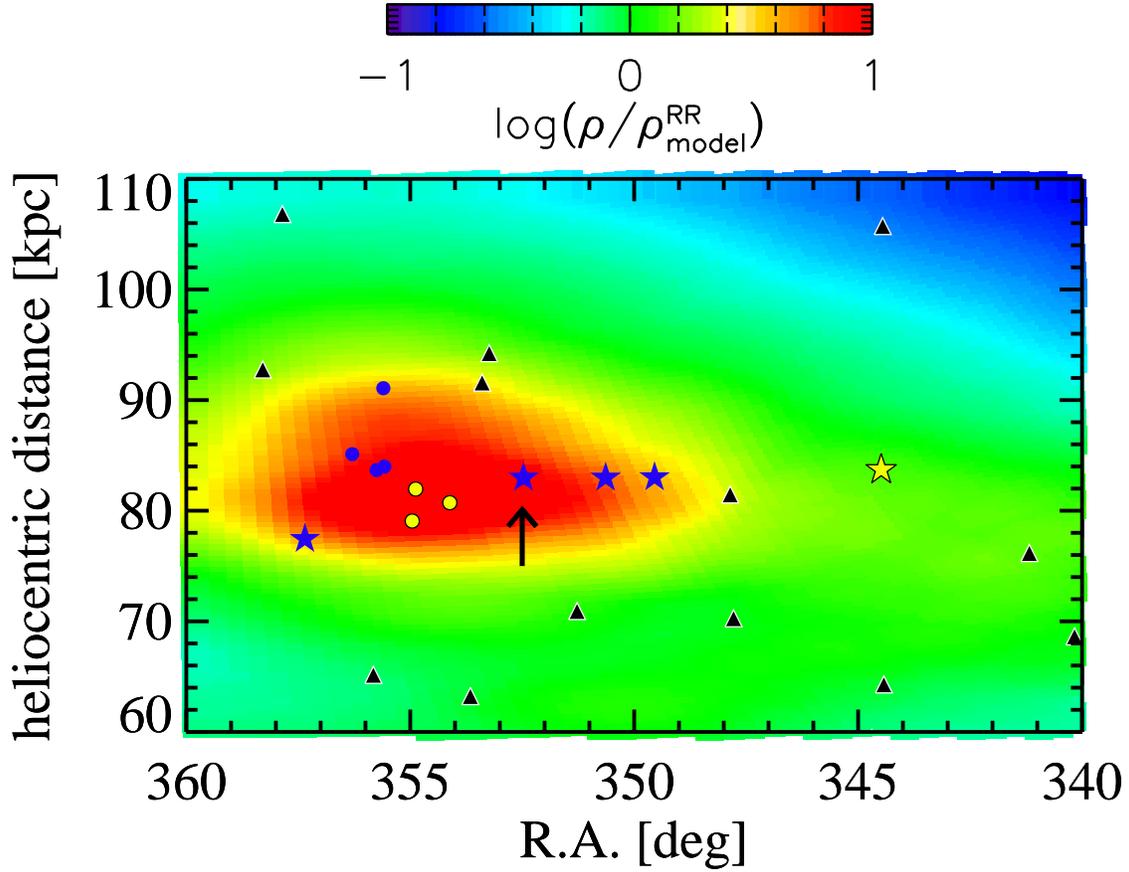}
\caption{
A zoom-in of Fig.~11 from \citet{sesar10} showing the spatial distribution of
RRab stars located in the vicinity of the Pisces overdensity (red and yellow
region). Due to their angular position, the heliocentric and galactocentric
distances of these stars are virtually the same. The colored background shows
the observed number density of RR Lyrae stars ($\rho$) relative to expected
smooth model number density ($\rho^{RR}_{model}$, Eq.~16 in \citealt{sesar10}).
The observed number density in the red region is 10 times greater than the model
prediction. The RRab stars were selected from the \citet{sesar10} stripe 82
catalog and have $|Dec|<1.25\arcdeg$. The stars spectroscopically observed by
\citet{kollmeier09} are shown as solid circles, those observed in this work are
shown as starred symbols, and stars not yet spectroscopically observed are shown
as triangles. The arrow points to a star that was observed in this work and by
\citet{kollmeier09}. The blue and yellow symbols show stars associated with
$\langle V_{gsr}\rangle=50$ km s$^{-1}$ and $\langle V_{gsr}\rangle=-52$ km
s$^{-1}$ velocity groups, respectively. The positions, velocities, and
metallicities of spectroscopically observed stars are listed in
Table~\ref{tab-data}.
\label{fig1}} 
\end{figure}

In Figure~\ref{fig1} we show the spatial distribution of $ab$-type RR Lyrae
(RRab) stars in the Pisces region. The stars were selected from the publicly
available \citet{sesar10} catalog. \citet{kollmeier09} obtained spectra of 8 of
these stars located in the densest part of this region (solid circles and one
starred symbol in Figure~\ref{fig1}). Our targets (starred symbols) for
spectroscopic observations were mainly chosen to expand the coverage in
direction of equatorial right ascension. As seen in Figure~\ref{fig1}, with the
addition of our targets all RR Lyrae stars in the densest part of the
overdensity now have spectroscopic measurements.

\begin{figure}
\epsscale{1.0}
\plotone{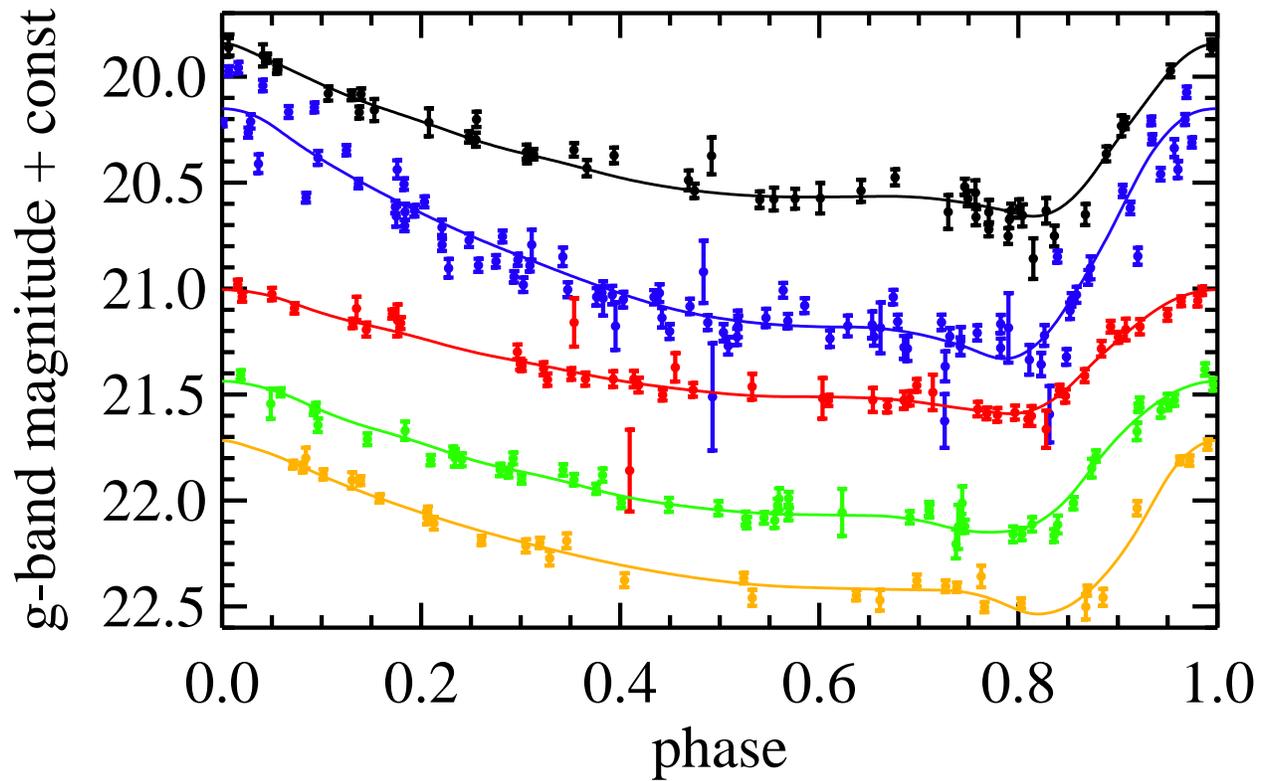}
\caption{
The $g$-band light curves of observed stripe 82 RR Lyrae stars, ordered as in
Table~\ref{tab-data}. The magnitudes are corrected for ithe ISM extinction using
the map from \citet{schlegel98}. The light curves are offset for clarity. The
solid lines show best-fit $g$-band templates from \citet{sesar10}.
\label{fig2}} 
\end{figure}

The positions, periods, and epochs of maximum brightness (HJD$_0$) of RR Lyrae
stars observed in this work are listed in Table~\ref{tab-data} and their
$g$-band light curves are shown in Figure~\ref{fig2}. The periods and epochs of
maximum brightness are precise to within a few seconds and minutes,
respectively, allowing the phase of pulsation to be determined with a 1\%
uncertainty \citep{sesar10}. The precise estimate of phase is important when
subtracting the velocity due to pulsations from the measured radial velocity, as
we explain in Section~\ref{spectro_obs}. Star 343892 (indicated by an arrow in
Figure~\ref{fig1}) is in common with \citeauthor{kollmeier09}'s sample and
provides a good check on the consistency of both works.

The heliocentric distances of RR Lyrae stars, $d$, are estimated from
measurements of their flux-averaged and dereddened $V$-band magnitudes,
$\langle V \rangle$ (for details see \citealt{sesar10}), and by assuming an
average value for their absolute magnitudes of $M_V=0.6$. This value was
obtained by using the \citet{chaboyer99} $M_V-[Fe/H]$ relation
\begin{equation}
M_V = (0.23\pm0.04)[Fe/H] + (0.93\pm0.12)\label{abs_mag}
\end{equation}
and adopting $[Fe/H]=-1.5$ as the metallicity of RRab stars (the median halo
metallicity, see \citealt{ivezic08}). This metallicity is used when calculating
distances even though for some stars we measure spectroscopic metallicities
later in the paper. The fractional error in the heliocentric distance due to
halo metallicity dispersion ($\sigma_{[Fe/H]}=0.3$ dex, \citealt{ivezic08}) and
RR Lyrae evolution off the zero-age horizontal branch, is estimated at 6\%
($\sim5$ kpc error in distance at 80 kpc). Due to their angular position, the
heliocentric and galactocentric distances of these stars are virtually the same.
The apparent $\langle V \rangle$ magnitudes of the observed stars range from
20.05 mag to 20.21 mag, corresponding to distances of 78 kpc and 84 kpc,
respectively. The spread in distances is consistent with the uncertainty in
distance.

\subsection{Spectroscopic Observations\label{spectro_obs}}

The spectroscopic observations were obtained during 2009 Aug 19-21 using the
Gemini Multi-object Spectrograph (GMOS; \citealt{Allington-Smith02}) mounted on
the 8.1-m Gemini South telescope at Cerro Pachon, Chile. Bad weather on two of
the three nights did not allow us to observe more than 5 targets. A $1\arcsec$
long-slit was used with the B1200 grating, resulting in a resolution of
$R=1820$, or $\sim2.5$ \AA~at 4500 \AA. This is equivalent to a resolution of
$\sim 150$ {\kms } at H$\beta$. The spectral range extends from 3800 {\AA } to
5250 \AA. The Pisces overdensity targets were observed in two sets of
2700s-exposures centered on $\lambda_{\rm central}=4500$ \AA~and
$\lambda_{\rm central}=4550$ \AA~(to account for the gaps in the detector), with
each exposure set bracketed by observations of a CuAr arc lamp. The difference
in resolution between the two setups (different $\lambda_{\rm central}$) is very
small (2.45 \AA~vs~2.47 \AA). Nevertheless, each set of spectra was reduced and
calibrated independently, and only then they were combined. The total exposure
time (1.5 hours) is $\lesssim 12\%$ of the pulsation cycle of these RR Lyrae
stars. Therefore, we do not expect excessive broadening of spectral features due
to the changing radial velocity during the pulsation cycle
(broadening $\lesssim 13$ \kms).

\begin{figure}
\epsscale{0.9}
\plotone{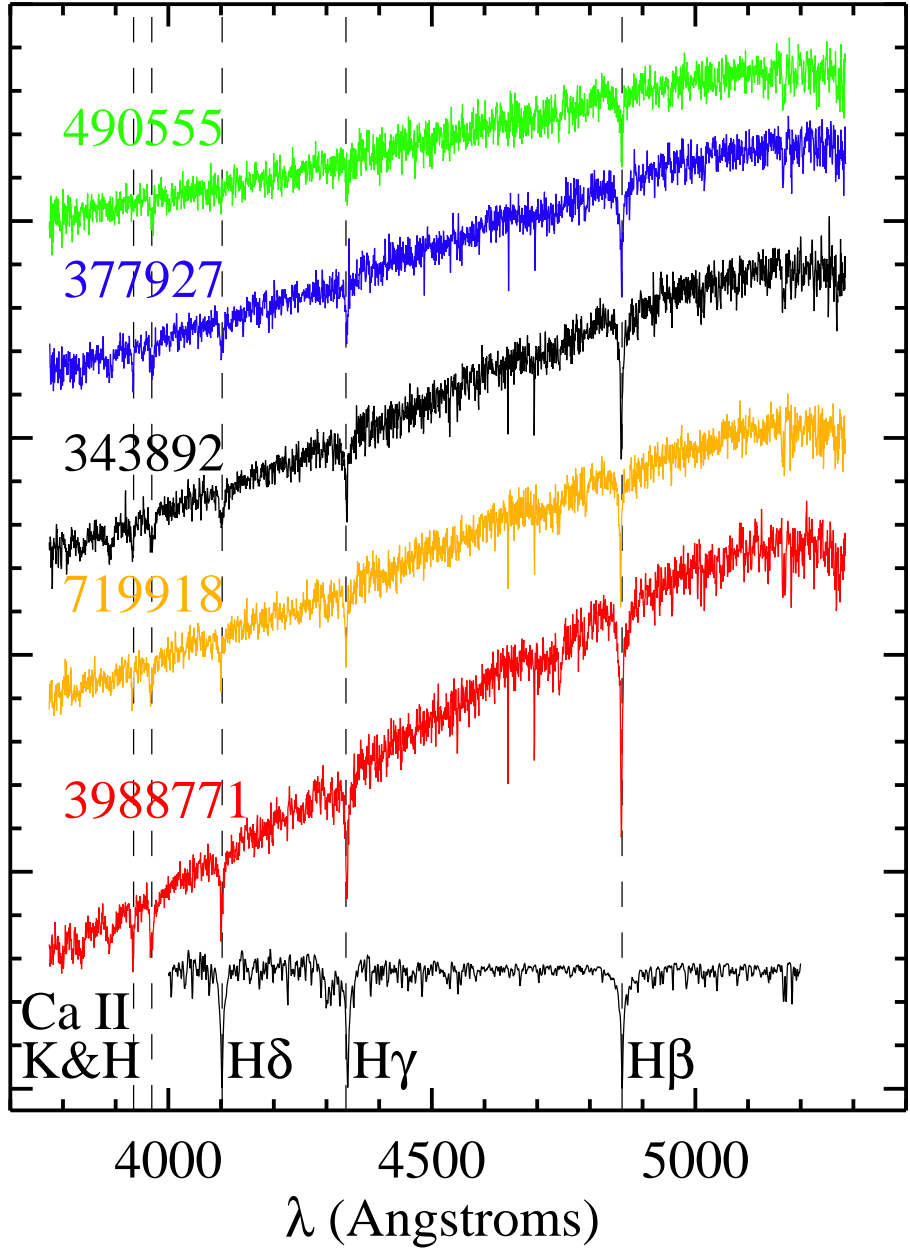}
\caption{
GMOS-S spectra of the 5 RR Lyrae stars observed in the Pisces overdensity,
ordered from lowest to highest S/N (top to bottom). An ELODIE template spectrum
(normalized to continuum) is shown at the bottom of the panel. The dashed lines
show the positions of important spectral features.
\label{fig3}} 
\end{figure}

Images were processed using the {\em gmos} package in IRAF. Cosmic rays were
removed using the L.A.~Cosmic routine\footnote{available at
\url[HREF]{http://www.astro.yale.edu/dokkum/lacosmic/}} \citep{dokum01}.
Wavelength calibration was done using $\sim 40$ CuAr arc spectral features over
the whole spectrum. The median root-mean-square (rms) scatter of the fit of the
wavelength calibration was 0.03 \AA, or 2 \kms. The signal-to-noise ratio (S/N)
of the final spectra ranged from 20 to 30 at $\sim 4000$ \AA. Figure~\ref{fig3}
shows the spectra of the 5 targets, ordered from the lowest to highest S/N.
Balmer lines (starting at H$\beta$) and Ca II H and K lines are easily seen in
the spectra.

We observed stars between phases 0.2 and 0.7 of the pulsation cycle in order to
avoid the discontinuity in the radial velocity curve near the maximum light. As
expected, the best S/N was achieved for stars observed during the descending
part of the light curve (phases 0.2 to 0.5) rather than during minimum light
(phase $>0.5$, e.g.~star 490555). Radial velocities, $V_r$, were obtained by
cross-correlating our spectra with 13 standard star spectra taken from the
ELODIE catalog\footnote{\url[HREF]{http://atlas.obs-hp.fr/elodie/}}
\citep{moultaka04}, which were degraded to the resolution of our observations
(convolved with the instrumental line profile and repixelized). For the
instrumental line profile, we use the line profile measured around the central
part of the spectrum (i.e., the line profile is assumed to be independent of
wavelength). The velocity measurements are not significantly affected by this
assumption, as measured velocities of three HD radial velocity standard stars
indicate below. The ELODIE radial velocity standard stars have spectral types
between F2 and F9. The systemic (center-of-mass) velocity of RR Lyrae stars,
$V_\gamma$, was obtained following the procedure described in detail in
\citet{vivas05,vivas08}. Here, we only summarize the most important points. We
used the radial velocity curve of RRab star X Arietis as a template. The
template was shifted in velocity to match the measurement for each star at the
corresponding phase and the systemic velocity is that of the shifted template at
phase 0.5.

Table~\ref{tab-error} shows different errors that contribute to the final
(systemic) velocity error. First, there is the cross-correlation error reported
by {\em fxcor}. This error includes the uncertainties due to broadening of
spectral features during an exposure, random errors in wavelength calibration,
and errors due to ELODIE template mismatch. The $\sigma_{cc}$ value listed in
Table~\ref{tab-error} is the average cross-correlation error value obtained from
13 cross-correlations. The radial velocity ($V_r$) listed in
Table~\ref{tab-data} is the mean of 13 $V_r$ values obtained from
cross-correlations, where each $V_r$ value is weighted by its cross-correlation
error. Therefore, radial velocities from better template fits contribute more
towards the final radial velocity value. The consistency of radial velocity
measurement obtained from cross-correlations is good ($\sim3$ \kms), as
indicated by the standard deviation $\sigma_{\rm templates}$.

The cross-correlation error may not fully represent the true error in the radial
velocity since it does not include systematic errors in the wavelength
calibration. To assess this issue, we cross-correlated 7 spectra of three radial
velocity standards (HD 154417, HD 155967 and HD 180482) which were observed
during our observing run with the same instrumental setup, with ELODIE
standards. The velocities of standards were recovered with a mean difference of
$-1.7$ {\kms } from literature values; the standard deviation of the differences
is only 3 \kms. The maximum difference is 6 \kms, which we assume as the
maximum error in the wavelength calibration. The error in the radial velocity
($\sigma_r$) was obtained by adding this value in quadrature with the
cross-correlation error. The average error of our radial velocity measurements
is 11 \kms. Finally, the error calculation for the systemic velocity includes
terms that take into account errors in the model radial velocity curve.
Following \citet{vivas05}, we use the following expresion in which the second
term accounts for likely uncertainties in the phase where the velocity curve
passes through the systemic velocity, and the third term is related to possible
variations in the slope of the velocity curve:
\begin{equation}
\sigma_\gamma^2=\sigma_r^2 + (119.5*0.1)^2 + (23.9*\Delta\phi)^2
\label{eq-sigma}
\end{equation}
where $\Delta\phi = \phi_{obs} - 0.5$, and $\phi_{obs}$ is the phase of
observation. The contributions of model uncertainties, $\sigma_{\rm model}$, to
final velocities are indicated in Table~\ref{tab-error}
($\sigma_{\rm model}^2 = (119.5*0.1)^2 + (23.9*\Delta\phi)^2$). These are of
the same order as the observed radial velocity errors. Therefore, to reduce the
uncertainty in the systemic velocity we need to reduce the uncertainty
introduced by the model, that is, we need to observe the radial velocity curve
more than once. The final errors in systemic velocities, $\sigma_\gamma$, range
from 10 to 18 \kms.

\begin{figure}
\epsscale{1.0}
\plotone{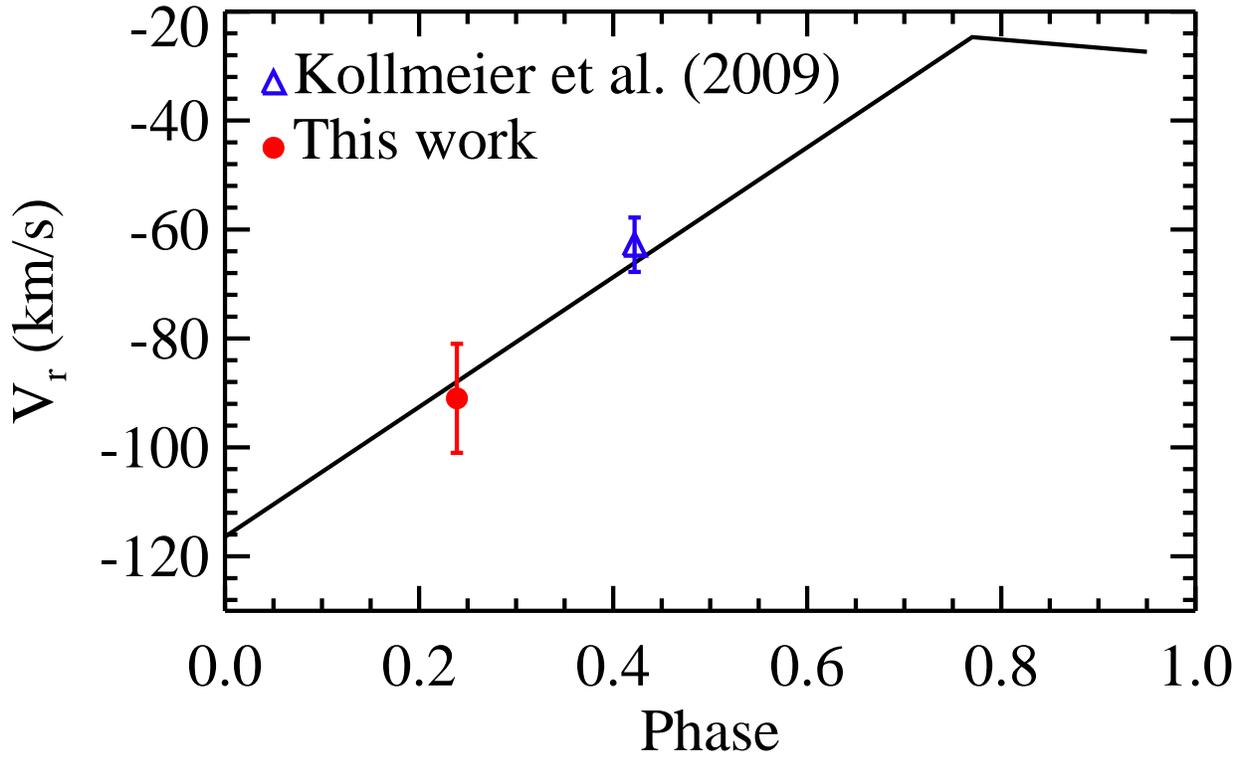}
\caption{
Radial velocity observations of star 343892 (indicated by an arrow in
Fig.~\ref{fig1}). The solid line is the template of the radial velocity curve of
RRab star X Arietis shifted in velocity to fit the observations.
\label{fig4}} 
\end{figure}

In the case of star 343892 we fitted the template to both our measurement and
the one reported by \citet{kollmeier09}. The fit is shown in Figure~\ref{fig4}.
The radial velocity curve of X Arietis fits well to both observations. This good
agreement gives us confidence that there are no large systematic differences
between our measurements and those reported by \citet{kollmeier09}. The systemic
velocity of this star reported in Table~\ref{tab-data} corresponds to the one
obtained by the fit of the template and differs somewhat from the value assumed
by \citeauthor{kollmeier09} (see Section~\ref{sec-koll}). In this case, the
final error was calculated as
$(\sigma_{\gamma,Koll}^{-2}+\sigma_{\gamma,This\, work}^{-2})^{-1/2}$ (standard
deviation of a weighted mean).

Finally, we calculated the velocity seen by an observer at the Sun who is at
rest with the Galactic center, $V_{gsr}$, by assuming solar motion of
$(v_U, v_V, v_W) = (10.0, 5.2, 7.2)$ {\kms } and $v_{LSR}=220$ {\kms }
\citep{binney98}.

The spectroscopic metallicities were measured following the method and
calibration of \citet{layden94} which involves plotting the pseudo-equivalent
width of Ca II K line, W(K), corrected for interstellar Ca absorption, against
the mean pseudo-equivalent widths of $\beta$, $\gamma$, and $\delta$ Balmer
lines, W(H) \citep[see][for details]{vivas05,vivas08}. With this method, we
estimated metallicity errors of 0.15 dex. Due to low S/N, the metallicity
for star 377927 has higher uncertainty (0.2 dex), and for star 490555 the
metallicity was not measured at all (see Table~\ref{tab-data}).

\subsection{Data from \citet{kollmeier09} \label{sec-koll}}

As shown in Figure~\ref{fig4}, there is a consistency between our radial
velocities and the ones measured by \citet{kollmeier09}. However, the methods
that were used to obtain the systemic velocity differ between the two works. In
order to avoid differences when combining samples, we decided to take their
reported radial velocities (uncorrected for pulsations) and use our method (that
is, fitting the template of X Arietis) to uniformly derive systemic velocities.
When our method is applied to their measurements, we find a systematic
difference of $\sim 15$ \kms (our values are larger than Kollmeier et al.'s).
Our main conclusions are not affected by this difference. For reference,
the original and revised \citet{kollmeier09} velocities are listed in
Table~\ref{tab-koll}. In the following section we use only the revised systemic
velocities.

Kollmeier et al.~also provide spectroscopic metallicities calibrated using the
\citet{gratton04} method. For star 343892 that is common to both samples, we
measure the metallicity of -1.1 dex while Kollmeier et al.~measured -1.2 dex.
It is quite encouraging that the two metallicity estimates are consistent within
uncertainties, considering different instruments and calibration methods. Given
this consistency, we simply combine our and Kollmeier et al.'s metallicities
when presenting results in the next section.

\section{Results and Discussion}

In this Section we address the following questions:
\begin{enumerate}
\item What is the mean velocity and velocity dispersion of our sample, and how
do these values compare to the literature?
\item Is the observed distribution of velocities a Gaussian or a non-Gaussian
distribution?
\item Are the observed velocity groups bound or unbound systems?
\item What does the metallicity of stars say about the progenitor(s) of the
streams?
\end{enumerate}

\begin{figure}
\epsscale{1.0}
\plotone{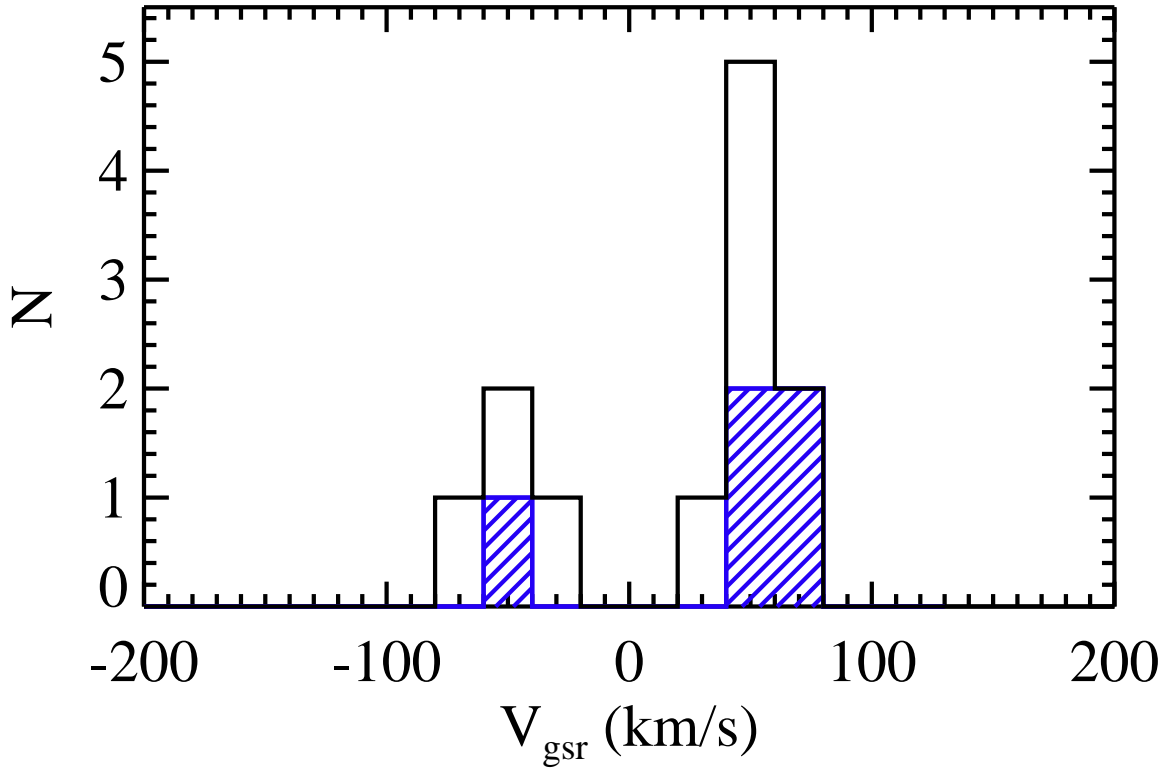}
\caption{
Histogram of $V_{gsr}$ of the combined sample of RR Lyrae stars in the Pisces
overdensity. The shaded histogram includes only stars observed in this work. The
bin size is 20 km s$^{-1}$, and is slightly larger than the largest radial
velocity error (18 km s$^{-1}$). The distribution of observed velocities seems
to be bimodal, with two velocity peaks centered on $\langle V_{gsr}\rangle=-52$
{\kms } and $\langle V_{gsr}\rangle=50$ {\kms }, respectively. See the text for
a discussion of statistical significance of this bimodality.
\label{fig5}} 
\end{figure}

Figure~\ref{fig5} shows the distribution of velocities of 12 RRab stars in the
Pisces overdensity from the combined data set. The weighted mean velocity,
velocity dispersion, and their uncertainties for the full sample are
$\langle V_{gsr} \rangle = 7\pm16$ km s$^{-1}$ and $\sigma_{gsr}=56\pm12$ km
s$^{-1}$. If the sample is divided into positive (8 stars) and negative
velocities (4 stars), the values are $\langle V^{pos}_{gsr} \rangle = 50\pm3$ km
s$^{-1}$, $\sigma^{pos}_{gsr}=10\pm3$ km s$^{-1}$,
$\langle V^{neg}_{gsr} \rangle = -52\pm11$ km s$^{-1}$, and
$\sigma^{neg}_{gsr}=23\pm9$ km s$^{-1}$.

The measured velocity dispersion for the full sample is consistent with the halo
velocity dispersion profile obtained by \cite{brown10} using a sample of 910
distant halo stars. By extrapolating their Equation 6 to 83 kpc, we obtain a
velocity dispersion of 73 km s$^{-1}$; a slightly higher dispersion than the one
we measure, but still within $\sim1\sigma$ uncertainties. Since the
\citep{brown10} sample covers a greater area of the sky than our sample (7300
deg$^2$ vs.~23 deg$^2$), it is also more likely to contain a greater number of
velocity groups at large distances than our sample. If these groups have
different internal kinematics (mean velocities and velocity dispersion), the net
effect will be a higher velocity dispersion in the sample with more velocity
groups. We have also searched the \citet{brown10} catalog for blue horizontal
branch stars that might be associated with the Pisces overdensity and found one
possible candidate. This star has R.A.~$=$355.575542 deg, Dec$=$0.326697 deg,
$V_{gsr}=101\pm17$ km s$^{-1}$, and is at a distance of 74 kpc. The uncertainty
in distance is $\sim6\%$, same as for RR Lyrae stars. However, we do not think 
this star is related to the Pisces overdensity as its velocity is at least
$3\sigma$ away from the positive velocity peak at
$\langle V^{pos}_{gsr} \rangle = 50$ \kms, where
$\sigma=\sigma^{pos}_{gsr}=10\pm3$ km s$^{-1}$.

The question of whether the distribution of velocities shown in
Figure~\ref{fig5} is a Gaussian or non-Gaussian distribution is an important
one, as deviations from normality are usually interpreted as a signature of
velocity groups in the smooth halo component \citep{harding01,duffau06,vivas08}.
To test for the presence of velocity groups in the halo, \citet{harding01}
recommend the \citet{shapiro65} (SW) statistical test of normality to be applied
to velocity histograms. This test is sensitive to many different deviations from
the Gaussian shape and does not depend on the choice of mean or dispersion of
the normal distribution (see \citealt{harding01} Section 5.1 and \citealt{ds86}
for more details on the test).

The null hypothesis of the SW test is that a sample came from a normally
distributed population. To test a sample, we use the {\em shapiro.test} routine
provided by the $R$ statistical
package\footnote{\url[HREF]{http://www.r-project.org/}}. This routine evaluates
the $W$ test statistic for a sample, and estimates its $P$-value for any $n$ in
the range $3\leqslant n \leqslant5000$, where $n$ is the sample size. For
example, if the $P$-value of a tested sample is less than 0.05, it can be
concluded that the data are not drawn from a normally distributed population at
the 95\% confidence level.

The $P$-value is the probability of obtaining a test statistic at least as
extreme as the one that was actually observed, assuming that the null hypothesis
is true. If we know that the null hypothesis is true, e.g.~we generate samples
by drawing values from a Gaussian distribution, the test is reliable if the
fraction of samples with $P<\alpha$ does not exceed $\alpha$. We use Monte Carlo
simulations to estimate the reliability of the $P$-value provided by the SW test
for small samples. We generate 10,000 samples and for each sample draw 12 values
from a Gaussian distribution. Assuming a fiducial value of $\alpha=0.01$, we
find that less than 1\% of samples have $P$-values less than 0.01. A similar
result is obtained for $\alpha=0.05$. These results show that the $P$-value
provided by the SW test is reliable even when testing small samples: the false
rejection rate is less than the adopted value of $\alpha$.

The SW test accepts single-valued data points and does not take into account
uncertainties in measurements. To account for uncertainties in measured
velocities when applying the SW test, we use Monte Carlo simulations. We
generate 10,000 samples and for each sample draw 12 velocities from 12 Gaussian
error distribution (one velocity per Gaussian). These Gaussians are centered on
measured velocities and have widths equal to uncertainties in measured
velocities. We then obtain a $P$-value for each sample using the SW test. The
most likely (mode) $P$-value for these samples is $\sim0.01$, the average
$P$-value is $\sim0.04$, and 95\% of samples have the $P$-value less than 0.12.
This analysis shows that there is only 4\% probability that the observed sample
was drawn from a Gaussian distribution.

The SW test does not have the ability to assess the rather peculiar nature of
the {\em shape} of the observed distribution (i.e., it cannot distinguish the
particular nature of non-Gaussianity). To include this information, we consider
the small chance (4\%) that the measured velocities were drawn from a Gaussian
distribution, and estimate the probability that it would produce the observed
distribution seen in Figure~\ref{fig5}. We generate 10,000 samples and for each
sample draw 12 velocities from a Gaussian centered on 7 km s$^{-1}$ and 56 km
s$^{-1}$ wide. Each sample is then divided into velocities greater and smaller
than 7 km s$^{-1}$, and dispersions are calculated for each group
($\sigma^{pos}$ and $\sigma^{neg}$). In only $\sim3\%$ of generated samples, the
positive velocities have $7<\sigma^{pos}<13$ and negative velocities have
$14<\sigma^{neg}<32$, where the selected ranges reflect the uncertainties in
velocity dispersions measured for the observed sample. Therefore, even if the
observed velocities were drawn from a Gaussian distribution, it is unlikely they
would create the distribution seen in Figure~\ref{fig5}.

In light of these results, we conclude that the hypothesis of normality can be
rejected at the $>95\%$ confidence level. The next simplest explanation for the
observed distribution of velocities is that we are observing two velocity
groups, one approaching at $\langle V^{neg}_{gsr} \rangle = -52\pm11$ km
s$^{-1}$, and the other one receding at $\langle V^{pos}_{gsr} \rangle = 50\pm3$
km s$^{-1}$.

In Figure~\ref{fig1}, the stars belonging to two velocity groups are shown in
different colors (blue for the positive velocity group and yellow for the
negative velocity group). One of the goals of this work was to probe the
extension of these groups in space. Our data show that both groups are quite
extended in space. The positive velocity group extends from
$349\degr < R.A. < 357\degr$, while the negative velocity group is mostly
centered around $R.A\sim355\degr$ with one star $10\degr$ away, at
$R.A\sim344\degr$. At 83 kpc, these angular extensions correspond to 9 and 15
kpc respectively. The large extensions of these groups are hard to reconcile
with bound systems. These groups are therefore most likely remnants of disrupted
dwarf galaxies, such as tidal streams, passing through the stripe 82 plane of
observation (similarly to the Sagittarius dSph trailing stream, see Figure 12 in
\citealt{sesar10} for an illustration).

In their work, \citet{kollmeier09} have observed two velocity groups but did not
consider the negative velocity group as a real system due to a small number of
stars (only three stars). We have repeated our statistical tests using their
sample and original velocities, and have found strong evidence of
non-Gaussianity in their distribution of velocities. The average $P$-value for
their sample is 0.03, and the probability of recreating a distribution of
velocities with two peaks from a single Gaussian distribution is less than
0.03\%. Therefore, we believe there was ample evidence in \citet{kollmeier09}
for detection of two velocity groups, even before four new observations from
this work are added. However, the real benefit of new observations, and of this
work, is the increased angular coverage of the Pisces region. The angular
extension of velocity groups is now more evident than in \citet{kollmeier09},
making the velocity groups more likely to be unbound (tidal streams) than bound
(globular clusters or dwarf galaxies).

By combining our and Kollmeier et al.~spectroscopic metallicities, we find that
the positive velocity group has a median metallicity of $[Fe/H]=-1.4$ dex and a
dispersion of 0.2 dex. The metallicity values for this group range from $-1.1$
dex to $-1.6$ dex. The metallicity of star 377927 was not used in this estimate
due to low S/N of its observed spectrum. The negative velocity group seems to be
more metal-poor, with a median metallicity of $[Fe/H]=-1.7$ dex, dispersion of
0.3 dex, and with metallicities ranging from $-1.6$ dex to $-1.9$ dex. The
number of stars in each group is small enough, and the errors are large enough,
that we cannot discard the hypothesis that the metallicities are drawn from the
same parent population. However, both groups are metal-poor suggesting that the
progenitors of these two substructures are systems with old, metal-poor
populations.

\section{Conclusions}

We confirm the existence of two kinematic groups in the Pisces overdensity,
located at 83 kpc from the Sun, by combining new spectroscopic observations with
those previously obtained by \citet{kollmeier09}. The spatial extent of groups
in the right ascension direction suggests that these are not bound systems but
rather debris of disrupted dwarf galaxies, most likely tidal streams. If these
are tidal streams, the question remains whether they have the same or different
progenitors. The first case (same progenitor) is not an improbable one. The
Sagittarius dSph galaxy has already shown us examples where debris stripped in
different epochs overlap in the same region of the sky (see Figure 1 in
\citealt{law05}). In such regions large gradients in $V_{gsr}$ are evident
along the same line of sight as some streams recede while others approach. The
second case (different progenitors) is supported by state-of-the-art simulations
of galaxy formation. These simulations predict numerous substructures of
different morphologies (streams, shells, clouds) in halos of Milky Way-size
galaxies \citep{johnston08}. Such substructures may overlap and may be difficult
to separate without wide and deep sky coverage.

Ongoing and upcoming wide-area surveys, such as Pan-STARRS \citep{kaiser02} and
LSST \citep{ivezic08b}, will provide the depth and sky coverage needed to
separate different halo substructures in coordinate space. With single-epoch and
co-added observations, these surveys will be able to observe main sequence F
dwarfs up to distances of 100 kpc and beyond, and will trace halo substructures
with greater fidelity. In addition, detections of overdensities in main sequence
and RR Lyrae stars will allow robust measurements of metallicities using a
photometric method described in \citet{sesar10}.
 
\acknowledgments

This research was supported in part by the National Science Foundation under
Grant No. PHY05-51164. B.~S.~and \v{Z}.~I.~acknowledge support by NSF grants AST
61-5991 and AST 07-07901, and by NSF grant AST 05-51161 to LSST for design and
development activity. Based on observations obtained at the Gemini Observatory,
which is operated by the Association of Universities for Research in Astronomy,
Inc., under a cooperative agreement with the NSF on behalf of the Gemini
partnership: the National Science Foundation (United States), the Science and
Technology Facilities Council (United Kingdom), the National Research Council
(Canada), CONICYT (Chile), the Australian Research Council (Australia),
Minist\'erio da Ci\v{e}ncia e Tecnologia (Brazil) and Ministerio de Ciencia,
Tecnolog\'ia e Innovaci\'on Productiva (Argentina). We are grateful to the
support staff at Gemini Observatory for their help during both the Phase II
preparation and observations of our program. We thank particularly Rodrigo
Carrasco for his valuable suggestions.

\begin{deluxetable}{ccccccccccccc}
\rotate
\tabletypesize{\scriptsize}
\tablecolumns{13}
\tablewidth{0pc}
\tablecaption{RR Lyrae Targets}
\tablehead{
\colhead{ID$^a$} & \colhead{R.A.$^b$} & \colhead{Dec$^b$} &
\colhead{Period} & \colhead{d$^c$} &\colhead{HJD$_0^d$} & \colhead{HJD$_{\rm spectrum}^e$} & 
\colhead{Phase$^f$} & \colhead{$V_r^g$} & \colhead{$V_\gamma^h$} & \colhead{$\sigma_\gamma^h$} &
\colhead{$V_{gsr}^i$} & \colhead{[Fe/H]\tablenotemark{j}} \\
\colhead{} & \colhead{(deg)} & \colhead{(deg)} &
\colhead{(d)} & \colhead{(kpc)} & \colhead{(+2450000 d)} & \colhead{(+1450000 d)} & 
\colhead{} & \colhead{(\kms)} & \colhead{(\kms)} & \colhead{(\kms)} &
\colhead{(\kms)} & \colhead{}
}
\startdata
 343892 & 352.469872  & -1.171239 & 0.59731 & 83  & 3352.59737 & 5064.64036 & 0.239 & -91 & -57  & 10 &  58  & -1.1 \\
 377927 & 350.637200  & -1.045226 & 0.48092 & 83  & 3637.73839 & 5064.87133 & 0.514 & -53 & -55  & 18 &  63  & -1.7* \\
 490555 & 349.544995  &  0.856966 & 0.63091 & 83  & 3666.70100 & 5065.86722 & 0.689 & -49 & -72  & 18 &  53  & \nodata \\
 719918 & 344.484556  &  0.333060 & 0.62337 & 84  & 4388.77431 & 5064.79280 & 0.466 & -186 &-182 & 16 & -51  & -1.9 \\
3988771 & 357.351934  &  0.759800 & 0.60580 & 78  & 4412.73725 & 5064.72046 & 0.226 & -69 & -36  & 16 &  76  & -1.7 \\
\enddata
\tablenotetext{a}{RR Lyrae ID number from the \citet{sesar10} catalog}
\tablenotetext{b}{Equatorial J2000.0 right ascension and declination}
\tablenotetext{c}{Heliocentric/galactocentric distance}
\tablenotetext{d}{Heliocentric Julian Date of maximum brightness}
\tablenotetext{e}{Heliocentric Julian Date when the spectrum was taken}
\tablenotetext{f}{Pulsation phase when the spectrum was taken}
\tablenotetext{g}{Radial velocity (not corrected for pulsations)}
\tablenotetext{h}{Systemic (center-of-mass) velocity (corrected for pulsations) and its uncertainty}
\tablenotetext{i}{Galactocentric rest-frame velocity}
\tablenotetext{j}{Spectroscopic metallicity, * denotes uncertain values}
\label{tab-data}
\end{deluxetable}

\begin{deluxetable}{ccccc}
\tablecolumns{5}
\tablewidth{0pc}
\tablecaption{Contributions to the error in $V_\gamma$}
\tablehead{
\colhead{ID} & \colhead{$\sigma_{cc}^a$} & \colhead{$\sigma_{\rm templates}^b$}
& \colhead{$\sigma_r^c$} & \colhead{$\sigma_{\rm model}^d$} \\
\colhead{} & \colhead{(\kms)} & \colhead{(\kms)} &
\colhead{(\kms)} & \colhead{(\kms)} 
}
\startdata
 343892 &  7.4  & 2.3 & 9.5  & 13.5 \\
 377927 &  11.8 & 3.7 & 13.2 & 12.0 \\
 490555 &  11.9 & 3.2 & 13.3 & 12.8 \\
 719918 &  8.9  & 0.5 & 10.7 & 12.0 \\
3988771 &  6.4  & 2.9 & 8.7  & 13.6 \\
\enddata
\tablenotetext{a}{Average cross-correlation error}
\tablenotetext{b}{Standard deviation of $V_r$ values obtained from cross-correlations}
\tablenotetext{c}{Uncertainty in measured radial velocity}
\tablenotetext{d}{Uncertainty in model radial velocity}
\label{tab-error}
\end{deluxetable}

\clearpage

\begin{deluxetable}{ccccccccc}
\tabletypesize{\scriptsize}
\tablecolumns{9}
\tablewidth{0pc}
\tablecaption{Revision of velocities from \citet{kollmeier09}}
\tablehead{
\colhead{ID} & \colhead{R.A.} & \colhead{Dec} &
\colhead{Phase} & \colhead{$V_r$} & \colhead{$V\gamma$} &
\colhead{$V\gamma$} & Difference & \colhead{$V_{gsr}$} \\
\colhead{} & \colhead{} & \colhead{} &
\colhead{} & \colhead{(Observed)} & \colhead{(Kollmeier et al.)} &
\colhead{(This work)} & \colhead{} & \colhead{(This work)} \\
\colhead{} & \colhead{(deg)} & \colhead{(deg)} &
\colhead{} & \colhead{(\kms)} &  \colhead{(\kms)} &
\colhead{(\kms)} & \colhead{(\kms)} & \colhead{(\kms)}
}
\startdata
 343892 & 352.469872  & -1.171239 & 0.422 &  -63 &  -68 &  -57 & 11 &  58 \\
3115111 & 354.116700  & -0.384237 & 0.362 & -156 & -156 & -140 & 16 & -26 \\
349151  & 354.878980  & -0.157729 & 0.436 & -192 & -198 & -184 & 14 & -71 \\
359035  & 354.955497  & -0.276307 & 0.335 & -192 & -189 & -172 & 17 & -59 \\
3981578 & 355.577586  & -0.008393 & 0.388 &  -82 &  -84 &  -69 & 15 &  44 \\
3974293 & 355.600850  & -0.623494 & 0.404 &  -84 &  -87 &  -72 & 15 &  39 \\
3944324 & 355.750750  & -0.173178 & 0.425 &  -65 &  -70 &  -56 & 14 &  56 \\
3879827 & 356.294558  & -0.804956 & 0.315 &  -73 &  -68 &  -51 & 17 &  58 \\
\enddata
\label{tab-koll}
\end{deluxetable}

\clearpage

\clearpage

\begin{thebibliography}

\bibitem[Allington-Smith et al.(2002)]{Allington-Smith02}Allington-Smith,
J.~et al.~2002, \pasp, 114, 892

\bibitem[{{Belokurov} {et~al.}(2007{\natexlab{a}}){Belokurov}, {Evans}, {Bell},
  {Irwin}, {Hewett}, {Koposov}, {Rockosi}, {Gilmore}, {Zucker}, {Fellhauer},
  {Wilkinson}, {Bramich}, {Vidrih}, {Rix}, {Beers}, {Schneider}, {Barentine},
  {Brewington}, {Brinkmann}, {Harvanek}, {Krzesinski}, {Long}, {Pan},
  {Snedden}, {Malanushenko}, \& {Malanushenko}}]{belokurov07a}
{Belokurov}, V. {et~al.} 2007{\natexlab{a}}, \apjl, 657, L89

\bibitem[{{Belokurov} {et~al.}(2007{\natexlab{b}}){Belokurov}, {Evans},
  {Irwin}, {Lynden-Bell}, {Yanny}, {Vidrih}, {Gilmore}, {Seabroke}, {Zucker},
  {Wilkinson}, {Hewett}, {Bramich}, {Fellhauer}, {Newberg}, {Wyse}, {Beers},
  {Bell}, {Barentine}, {Brinkmann}, {Cole}, {Pan}, \& {York}}]{belokurov07b}
{Belokurov}, V. {et~al.} 2007{\natexlab{b}}, \apj, 658, 337

\bibitem[{{Binney} \& {Merrifield}(1998)}]{binney98}
{Binney}, J., \& {Merrifield}, M. 1998, {Galactic astronomy}, ed. {Binney,
  J.~\& Merrifield, M.}

\bibitem[Brown et al.(2010)]{brown10}Brown, W.~R.~et al.~2010, \aj, 139, 59

\bibitem[Chaboyer(1999)]{chaboyer99}Chaboyer, B. 1999, in ``Post-Hipparcos
cosmic candles'', Eds. A. Heck \& F. Caputo, Kluwer Academic Publishers, p. 111

\bibitem[{{Clewley} \& {Kinman}(2006)}]{clewley06}
{Clewley}, L., \& {Kinman}, T.~D. 2006, \mnras, 371, L11

\bibitem[D'Agostino \& Stephens(1986)]{ds86}D'Agostino, R. B. \& Stephens, M. A.
1986, Goodness of Fit Techniques (New York: Marcel Dekker)

\bibitem[{{Duffau} {et~al.}(2006){Duffau}, {Zinn}, {Vivas}, {Carraro},
  {M{\'e}ndez}, {Winnick}, \& {Gallart}}]{duffau06}
{Duffau}, S., {Zinn}, R., {Vivas}, A.~K., {Carraro}, G., {M{\'e}ndez}, R.~A.,
  {Winnick}, R., \& {Gallart}, C. 2006, \apjl, 636, L97

\bibitem[Gratton et al.(2004)]{gratton04}Gratton, R.~G.~et al. 2004, \aap, 421,
937

\bibitem[Grillmair(2009)]{grillmair09}Grillmair, C.~J. 2009, \apj, 693, 1118

\bibitem[Harding et al.(2001)]{harding01}Harding, P.~et al. 2001, \aj, 122, 1397
  
\bibitem[{{Helmi}(2008)}]{helmi08} {Helmi}, A. 2008, \aapr, 15, 145

\bibitem[Ibata et al.(2002)]{ibata02}Ibata, R.~A.~et al. 2002, \mnras, 332, 921

\bibitem[{{Ibata} {et~al.}(2003){Ibata}, {Irwin}, {Lewis}, {Ferguson}, \&
  {Tanvir}}]{ibata03}
{Ibata}, R.~A., {Irwin}, M.~J., {Lewis}, G.~F., {Ferguson}, A.~M.~N., \&
  {Tanvir}, N. 2003, \mnras, 340, L21

\bibitem[{{Ivezi{\'c}} {et~al.}(2000){Ivezi{\'c}}, {Goldston}, {Finlator},
  {Knapp}, {Yanny}, {McKay}, {Amrose}, {Krisciunas}, {Willman}, {Anderson},
  {Schaber}, {Erb}, {Logan}, {Stubbs}, {Chen}, {Neilsen}, {Uomoto}, {Pier},
  {Fan}, {Gunn}, {Lupton}, {Rockosi}, {Schlegel}, {Strauss}, {Annis},
  {Brinkmann}, {Csabai}, {Doi}, {Fukugita}, {Hennessy}, {Hindsley}, {Margon},
  {Munn}, {Newberg}, {Schneider}, {Smith}, {Szokoly}, {Thakar}, {Vogeley},
  {Waddell}, {Yasuda}, \& {York}}]{ivezic00}
{Ivezi{\'c}}, {\v Z}. {et~al.} 2000, \aj, 120, 963

\bibitem[{{Ivezi{\'c}} {et~al.}(2008a){Ivezi{\'c}}, {Sesar}, {Juri{\'c}},
  {Bond}, {Dalcanton}, {Rockosi}, {Yanny}, {Newberg}, {Beers}, {Allende
  Prieto}, {Wilhelm}, {Lee}, {Sivarani}, {Norris}, {Bailer-Jones}, {Re
  Fiorentin}, {Schlegel}, {Uomoto}, {Lupton}, {Knapp}, {Gunn}, {Covey},
  {Smith}, {Miknaitis}, {Doi}, {Tanaka}, {Fukugita}, {Kent}, {Finkbeiner},
  {Munn}, {Pier}, {Quinn}, {Hawley}, {Anderson}, {Kiuchi}, {Chen}, {Bushong},
  {Sohi}, {Haggard}, {Kimball}, {Barentine}, {Brewington}, {Harvanek},
  {Kleinman}, {Krzesinski}, {Long}, {Nitta}, {Snedden}, {Lee}, {Harris},
  {Brinkmann}, {Schneider}, \& {York}}]{ivezic08}
{Ivezi{\'c}}, {\v Z}. {et~al.}, 2008a, \apj, 684, 287

\bibitem[Ivezi\'c et al.(2008b)]{ivezic08b}Ivezi\'c, \v{Z}. et al. 2008b,
arXiv:0805.2366

\bibitem[Johnston et al.(2008)]{johnston08}Johnston, K.~V.~et al.~2008, \apj,
689, 936

\bibitem[{{Juri{\'c}} {et~al.}(2008){Juri{\'c}}, {Ivezi{\'c}}, {Brooks},
  {Lupton}, {Schlegel}, {Finkbeiner}, {Padmanabhan}, {Bond}, {Sesar},
  {Rockosi}, {Knapp}, {Gunn}, {Sumi}, {Schneider}, {Barentine}, {Brewington},
  {Brinkmann}, {Fukugita}, {Harvanek}, {Kleinman}, {Krzesinski}, {Long},
  {Neilsen}, {Nitta}, {Snedden}, \& {York}}]{juric08}
{Juri{\'c}}, M. {et~al.} 2008, \apj, 673, 864

\bibitem[Kaiser et al.(2002)]{kaiser02}Kaiser, N. et al. 2002, \procspie, 4836,
154

\bibitem[Klypin et al.(1999)]{klypin99}Klypin, A.~et al.~1999, \apj, 522, 82

\bibitem[{{Kollmeier} {et~al.}(2009){Kollmeier}, {Gould}, {Shectman},
  {Thompson}, {Preston}, {Simon}, {Crane}, {Ivezi{\'c}}, \&
  {Sesar}}]{kollmeier09}
{Kollmeier}, J.~A. {et~al.} 2009, \apjl, 705, L158

\bibitem[{{Law} {et~al.}(2005){Law}, {Johnston}, \& {Majewski}}]{law05}
{Law}, D.~R., {Johnston}, K.~V., \& {Majewski}, S.~R. 2005, \apj, 619, 807

\bibitem[{{Layden}(1994)}]{layden94}
{Layden}, A.~C. 1994, \aj, 108, 1016

\bibitem[{{Majewski} {et~al.}(2003){Majewski}, {Skrutskie}, {Weinberg}, \&
  {Ostheimer}}]{majewski03}
{Majewski}, S.~R., {Skrutskie}, M.~F., {Weinberg}, M.~D., \& {Ostheimer}, J.~C.
  2003, \apj, 599, 1082

\bibitem[{{Mateo}(1998)}]{mateo98}
{Mateo}, M.~L. 1998, \araa, 36, 435

\bibitem[{{Moultaka} {et~al.}(2004){Moultaka}, {Ilovaisky}, {Prugniel}, \&
  {Soubiran}}]{moultaka04}
{Moultaka}, J., {Ilovaisky}, S.~A., {Prugniel}, P., \& {Soubiran}, C. 2004,
  \pasp, 116, 693

\bibitem[{{Newberg} {et~al.}(2007){Newberg}, {Yanny}, {Cole}, {Beers}, {Re
  Fiorentin}, {Schneider}, \& {Wilhelm}}]{newberg07}
{Newberg}, H.~J., {Yanny}, B., {Cole}, N., {Beers}, T.~C., {Re Fiorentin}, P.,
  {Schneider}, D.~P., \& {Wilhelm}, R. 2007, \apj, 668, 221

\bibitem[{{Prior} {et~al.}(2009){Prior}, {Da Costa}, {Keller}, \&
  {Murphy}}]{prior09}
{Prior}, S.~L., {Da Costa}, G.~S., {Keller}, S.~C., \& {Murphy}, S.~J. 2009,
  \apj, 691, 306

\bibitem[{{Rocha-Pinto} {et~al.}(2004){Rocha-Pinto}, {Majewski}, {Skrutskie},
  {Crane}, \& {Patterson}}]{rocha04}
{Rocha-Pinto}, H.~J., {Majewski}, S.~R., {Skrutskie}, M.~F., {Crane}, J.~D., \&
  {Patterson}, R.~J. 2004, \apj, 615, 732

\bibitem[{{Schlegel} {et~al.}(1998){Schlegel}, {Finkbeiner}, \&
  {Davis}}]{schlegel98}
{Schlegel}, D.~J., {Finkbeiner}, D.~P., \& {Davis}, M. 1998, \apj, 500, 525

\bibitem[{{Sesar} {et~al.}(2007){Sesar}, {Ivezi{\'c}}, {Lupton}, {Juri{\'c}},
  {Gunn}, {Knapp}, {DeLee}, {Smith}, {Miknaitis}, {Lin}, {Tucker}, {Doi},
  {Tanaka}, {Fukugita}, {Holtzman}, {Kent}, {Yanny}, {Schlegel}, {Finkbeiner},
  {Padmanabhan}, {Rockosi}, {Bond}, {Lee}, {Stoughton}, {Jester}, {Harris},
  {Harding}, {Brinkmann}, {Schneider}, {York}, {Richmond}, \& {Vanden
  Berk}}]{sesar07}
{Sesar}, B. {et~al.} 2007, \aj, 134, 2236

\bibitem[{{Sesar} {et~al.}(2010){Sesar}, {Ivezi{\'c}}, {Grammer}, {Morgan},
  {Becker}, {Juri{\'c}}, {De Lee}, {Annis}, {Beers}, {Fan}, {Lupton}, {Gunn},
  {Knapp}, {Jiang}, {Jester}, {Johnston}, \& {Lampeitl}}]{sesar10}
{Sesar}, B. {et~al.} 2010, \apj, 708, 717

\bibitem[{{Shapiro} \& {Wilk}(1965)}]{shapiro65}
{Shapiro}, S.~S., \& {Wilk}, M.~B. 1965, Biometrika, 52, 591

\bibitem[{{Starkenburg} {et~al.}(2009){Starkenburg}, {Helmi}, {Morrison},
  {Harding}, {van Woerden}, {Mateo}, {Olszewski}, {Sivarani}, {Norris},
  {Freeman}, {Shectman}, {Dohm-Palmer}, {Frey}, \& {Oravetz}}]{starkenburg09}
{Starkenburg}, E. {et~al.} 2009, \apj, 698, 567

\bibitem[Schlaufman et al.(2009)]{schlaufman09}Schlaufman, K.~C. et al. 2009,
\apj, 703, 2177

\bibitem[van Dokum(2001)]{dokum01}van Dokum, P.~G.~2001, \pasp, 113, 1420

\bibitem[{{Vivas} {et~al.}(2008){Vivas}, {Jaff{\'e}}, {Zinn}, {Winnick},
  {Duffau}, \& {Mateu}}]{vivas08}
{Vivas}, A.~K., {Jaff{\'e}}, Y.~L., {Zinn}, R., {Winnick}, R., {Duffau}, S., \&
  {Mateu}, C. 2008, \aj, 136, 1645

\bibitem[{{Vivas} {et~al.}(2001){Vivas}, {Zinn}, {Andrews}, {Bailyn}, {Baltay},
  {Coppi}, {Ellman}, {Girard}, {Rabinowitz}, {Schaefer}, {Shin}, {Snyder},
  {Sofia}, {van Altena}, {Abad}, {Bongiovanni}, {Brice{\~n}o}, {Bruzual},
  {Della Prugna}, {Herrera}, {Magris}, {Mateu}, {Pacheco}, {S{\'a}nchez},
  {S{\'a}nchez}, {Schenner}, {Stock}, {Vicente}, {Vieira}, {Ferr{\'{\i}}n},
  {Hernandez}, {Gebhard}, {Honeycutt}, {Mufson}, {Musser}, \&
  {Rengstorf}}]{vivas01}
{Vivas}, A.~K. {et~al.} 2001, \apjl, 554, L33

\bibitem[{{Vivas} {et~al.}(2005){Vivas}, {Zinn}, \& {Gallart}}]{vivas05}
{Vivas}, A.~K., {Zinn}, R., \& {Gallart}, C. 2005, \aj, 129, 189

\bibitem[{{Watkins} {et~al.}(2009){Watkins}, {Evans}, {Belokurov}, {Smith},
  {Hewett}, {Bramich}, {Gilmore}, {Irwin}, {Vidrih}, {Wyrzykowski}, \&
  {Zucker}}]{watkins09}
{Watkins}, L.~L. {et~al.} 2009, \mnras, 398, 1757

\bibitem[{{Yanny} {et~al.}(2000){Yanny}, {Newberg}, {Kent},
  {Laurent-Muehleisen}, {Pier}, {Richards}, {Stoughton}, {Anderson}, {Annis},
  {Brinkmann}, {Chen}, {Csabai}, {Doi}, {Fukugita}, {Hennessy}, {Ivezi{\'c}},
  {Knapp}, {Lupton}, {Munn}, {Nash}, {Rockosi}, {Schneider}, {Smith}, \&
  {York}}]{yanny00}
{Yanny}, B. {et~al.} 2000, \apj, 540, 825

\bibitem[{{Yanny} {et~al.}(2003){Yanny}, {Newberg}, {Grebel}, {Kent},
  {Odenkirchen}, {Rockosi}, {Schlegel}, {Subbarao}, {Brinkmann}, {Fukugita},
  {Ivezic}, {Lamb}, {Schneider}, \& {York}}]{yanny03}
{Yanny}, B. {et~al.} 2003, \apj, 588, 824

\bibitem[{{York} {et~al.}(2000){York}, {Adelman}, {Anderson}, {Anderson},
  {Annis}, {Bahcall}, {Bakken}, {Barkhouser}, {Bastian}, {Berman}, {Boroski},
  {Bracker}, {Briegel}, {Briggs}, {Brinkmann}, {Brunner}, {Burles}, {Carey},
  {Carr}, {Castander}, {Chen}, {Colestock}, {Connolly}, {Crocker}, {Csabai},
  {Czarapata}, {Davis}, {Doi}, {Dombeck}, {Eisenstein}, {Ellman}, {Elms},
  {Evans}, {Fan}, {Federwitz}, {Fiscelli}, {Friedman}, {Frieman}, {Fukugita},
  {Gillespie}, {Gunn}, {Gurbani}, {de Haas}, {Haldeman}, {Harris}, {Hayes},
  {Heckman}, {Hennessy}, {Hindsley}, {Holm}, {Holmgren}, {Huang}, {Hull},
  {Husby}, {Ichikawa}, {Ichikawa}, {Ivezi{\'c}}, {Kent}, {Kim}, {Kinney},
  {Klaene}, {Kleinman}, {Kleinman}, {Knapp}, {Korienek}, {Kron}, {Kunszt},
  {Lamb}, {Lee}, {Leger}, {Limmongkol}, {Lindenmeyer}, {Long}, {Loomis},
  {Loveday}, {Lucinio}, {Lupton}, {MacKinnon}, {Mannery}, {Mantsch}, {Margon},
  {McGehee}, {McKay}, {Meiksin}, {Merelli}, {Monet}, {Munn}, {Narayanan},
  {Nash}, {Neilsen}, {Neswold}, {Newberg}, {Nichol}, {Nicinski}, {Nonino},
  {Okada}, {Okamura}, {Ostriker}, {Owen}, {Pauls}, {Peoples}, {Peterson},
  {Petravick}, {Pier}, {Pope}, {Pordes}, {Prosapio}, {Rechenmacher}, {Quinn},
  {Richards}, {Richmond}, {Rivetta}, {Rockosi}, {Ruthmansdorfer}, {Sandford},
  {Schlegel}, {Schneider}, {Sekiguchi}, {Sergey}, {Shimasaku}, {Siegmund},
  {Smee}, {Smith}, {Snedden}, {Stone}, {Stoughton}, {Strauss}, {Stubbs},
  {SubbaRao}, {Szalay}, {Szapudi}, {Szokoly}, {Thakar}, {Tremonti}, {Tucker},
  {Uomoto}, {Vanden Berk}, {Vogeley}, {Waddell}, {Wang}, {Watanabe},
  {Weinberg}, {Yanny}, \& {Yasuda}}]{york00}
{York}, D.~G. {et~al.} 2000, \aj, 120, 1579

\end{thebibliography}
\end{document}